\documentclass[runningheads]{llncs}
\usepackage{hyperref} 
\usepackage{subcaption}
\usepackage{comment}
\usepackage[T1]{fontenc}
\usepackage{graphicx}
\makeatletter
\def\blfootnote{\xdef\@thefnmark{}\@footnotetext}
\makeatother
\usepackage{fancyhdr}
\providecommand{\runtitle}{Undefined}
\providecommand{\runauthor}{Undefined}
\renewcommand{\runtitle}{AI-Assisted Music Production: A User Study on Text-to-Music Models}
\renewcommand{\runauthor}{F. Ronchini et al.}
\begin{document}
\title{AI-Assisted Music Production: A User Study on Text-to-Music Models}
\author{Francesca Ronchini\orcidID{0000-0001-6897-1645} \and
Luca Comanducci\orcidID{0000-0002-4167-5173} \and
Simone Marcucci\and
Fabio Antonacci\orcidID{0000-0003-4545-0315}}
\institute{Dipartimento di Elettronica, Informazione e Bioingegneria (DEIB), Politecnico di Milano
Piazza Leonardo Da Vinci 32, 20133 Milan, Italy\\\email{\{name.surname\}@polimi.it}}
\maketitle              
%
\blfootnote{\includegraphics[scale=0.25]{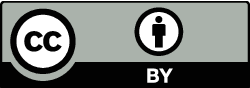} All rights remain with the authors under the Creative Commons Attribution 4.0 International License (CC BY 4.0).\\
Proc. of the 17th Int. Symposium on Computer Music Multidisciplinary Research,\\
London, United Kingdom, 2025}
\begin{abstract}
Text-to-music models have revolutionized the creative landscape, offering new possibilities for music creation. Yet their integration into musicians’ workflows remains underexplored. This paper presents a case study on how TTM models impact music production, based on a user study of their effect on producers' creative workflows. Participants produce tracks using a custom tool combining TTM and source separation models. Semi-structured interviews and thematic analysis reveal key challenges, opportunities, and ethical considerations. The findings offer insights into the transformative potential of TTMs in music production, as well as challenges in their real-world integration.

\end{abstract}
\section{Introduction}
\label{sec:intro}

Deep learning is rapidly advancing music generation, particularly through Text-to-Music (TTM) models that generate audio from text prompts~\cite{zhao2025artificial,ronchini2025mind}. These models are transforming music composition and production~\cite{kamath2024sound,ma2024foundation}, enabling creative experimentation even for those without formal training, as seen with Suno~\cite{suno2024} and Udio~\cite{udio2024}. Despite advancements in AI music generation, TTM models still struggle to interpret musicians’ controls~\cite{zang2024interpretation,ronchini2024paguri}, underscoring the need for focused research on creator interaction and collaboration. Ronchini et al.~\cite{ronchini2024paguri} conducted a user experience study exploring how musicians interact with a TTM model enhanced by fine-tuning techniques to adapt to users' specific preferences. However, the study was constrained by a homogeneous participant pool and a broad focus. Parallel to this paper, Fu et al.~\cite{fu2025exploring} conducted a similar investigation focused on music production, using a limited sample of university students with little to no production experience, where pressure due to deadlines and assignments may have influenced the results.

\begin{figure}[t!]
  \centering
  \includegraphics[alt={Workflow of the experimental procedure.},width=.9\columnwidth]{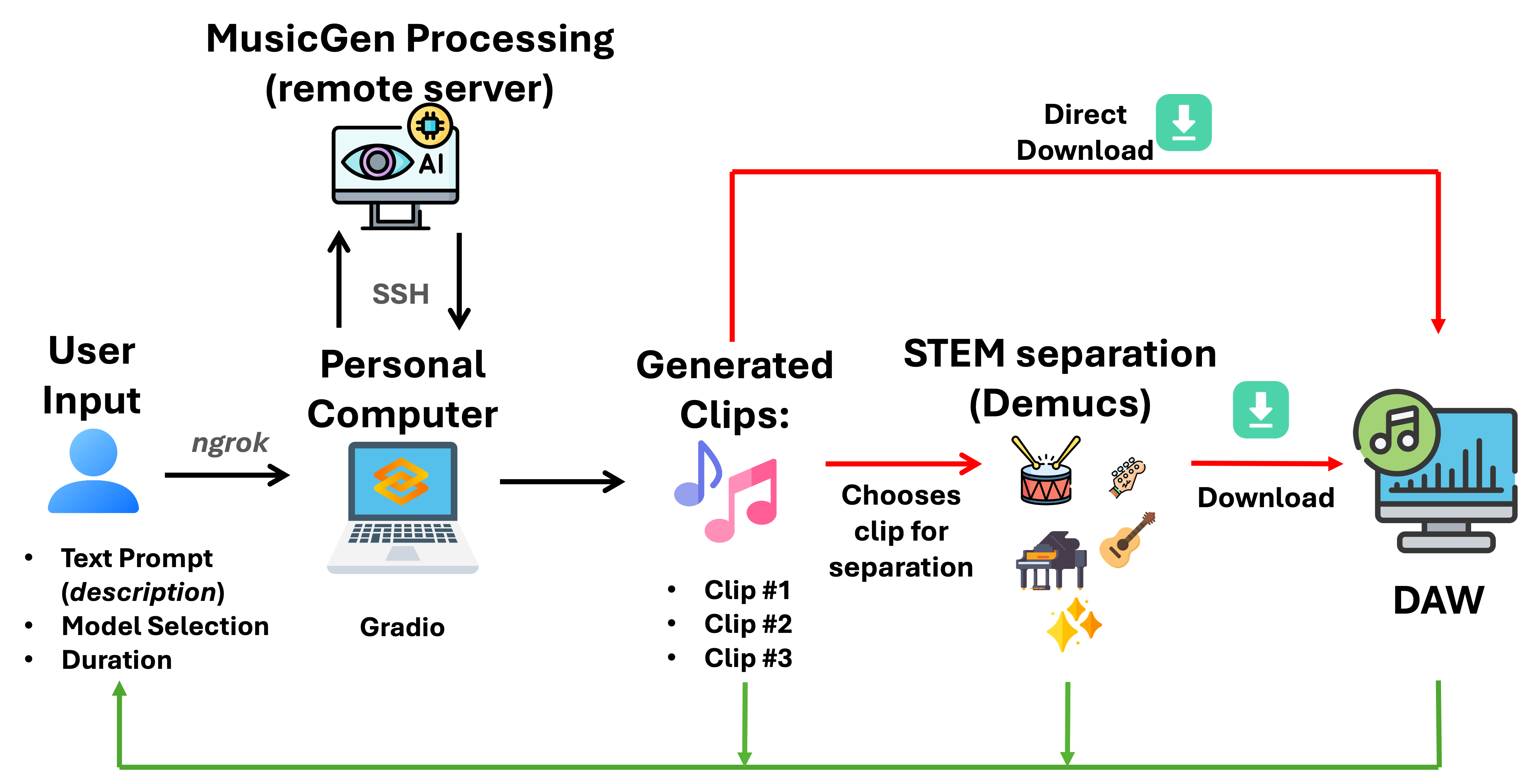}
  \caption{Workflow of the experimental procedure. The red lines mean optional actions, while the green lines mean the reiteration process.}
  \label{fig:example}
\end{figure} 
This paper overcomes previous limitations, proposing a study to better investigate the interpretation gap~\cite{zang2024interpretation} by conducting a user case study on AI-assisted music production. The research involves 17 music producers with varying levels of experience, and evaluate how TTM models influence productivity and effectiveness, the usability of AI tools as creative collaborators, and the overall perspective on incorporating TTM models into creative workflow. Participants were invited to use their preferred setups to produce a musical track with the help of a custom-developed tool that integrates a TTM model and an AI-based source separation model, providing producers with greater control and flexibility over the generated composition~\cite{rouard2025musicgen}. 
Fig.~\ref{fig:example} illustrates the workflow of the experiment.
We believe these results offer insightful perspectives on user experiences and ethical implications of TTM models. Supplementary material, including videos of interaction and code for replication, are available \href{https://lucacoma.github.io/AiMusicProductionUserStudy/}{online}.

\section{Background}
\label{sec:background}


\textbf{Text-to-music Models}. Several text-based generative models, based on different approaches, have been proposed for the raw audio and music generation task~\cite{zhao2025artificial}. The first model proposed is AudioLM~\cite{borsos2023audiolm}, 
which  was further developed in MusicLM~\cite{agostinelli2023musiclm}.  
Subsequently, transformer-based models have been proposed~\cite{kreuk2022audiogen,copet2023simple,ziv2024masked}, 
and diffusion-based models have gained attention~\cite{yang2023diffsound,liu2023audioldm2,evans2024stable}. 
Recently, commercial solutions like Suno AI~\cite{suno2024} and Udio~\cite{udio2024} have been introduced. 
However, we opted not to consider them in order to adhere to the principles of open science, enabling fellow researchers to reproduce our experiments. Moreover, the use of closed-source models would have prevented us from integrating the source separation technique into a single interface. In recent research, control inputs such as tempo, rhythm, chords, and melody have been integrated alongside text~\cite{wu2024music,lan2024musicongen,tal2024joint}. As text remains the primary control in these models, this study focuses on text-based approaches. For a more detailed survey, the reader is referred to~\cite{zhao2025artificial,ma2024foundation}.

\textbf{Music Source Separation Models}. Data-driven models for Music Source Separation are categorized into spectrogram-based~\cite{takahashi2020d3net} and waveform-based models~\cite{defossez2019demucs}. 
In~\cite{defossez2021hybrid}, the authors propose hybrid approaches that combine both temporal and spectral domains. Spleeter~\cite{hennequin2020spleeter} leverages pre-trained models. Recently, Hybrid Transformer Demucs (HT-Demucs)~\cite{rouard2023hybrid}, inspired by Hybrid Demucs~\cite{defossez2021hybrid}, and Banquet~\cite{watcharasupat2024stem} have been introduced. Most models separate audio into four stems: vocals, drums, bass, and other (VDBO). Spleeter adds piano for five stems, HT-Demucs includes guitar for six, and Banquet generates multiple stems~\cite{watcharasupat2024stem}.

\textbf{Co-creativity and Human-AI interaction}. Several studies explore Human-AI interaction in music generative models~\cite{morreale2016collaborating,louie2020novice}, the perception of AI-generated music~\cite{chu2022empirical,sarmento2024between}, and opportunities and challenges of AI for music~\cite{chu2022empirical,newman2023human}. In~\cite{louie2022expressive}, the authors evaluate generative models and AI-steering interfaces in downstream creative tasks. 
Zhou et al.~\cite{zhou2021interactive} explore ways to enhance user interaction by allowing direct control over a model’s sampling behavior, while Huang et al.~\cite{newman2023human} study human-AI co-creation, identifying challenges musicians face when composing with AI. 
Several interfaces have been proposed to improve interaction between generative models and users~\cite{bougueng2022calliope,louie2020novice,rau2022visualization,zhang2021cosmic,zhou2021interactive,yakura2023iteratta}, AI-assisted music composition~\cite{rau2025maico}, along with creative supporting tools designed to enhance collaborative creativity~\cite{chung2021intersection,frich2019mapping,kamath2024sound}. However, research of TTM integration into music creation workflows remains limited, but crucial~\cite{ronchini2024paguri,fu2025exploring,zang2024interpretation}.

\section{Study design and method}
\label{sec:method}

\textbf{Models and framework selection}. We chose MusicGen~\cite{copet2023simple} for music generation and HT Demucs~\cite{rouard2023hybrid} for source separation due to their performance trade-offs and easy integration into a unified framework, ensuring reproducibility. 
MusicGen is a text- or melody-conditioned autoregressive model available in three sizes, while HT Demucs~\cite{rouard2023hybrid} uses dual U-Nets and a Transformer to separate instruments. We use the \textit{htdemucs\_6s} model, 
which separates VDBO, guitar and piano. Combining music separation with generation enhances composition allows independent manipulation of individual sources. MusicGen-Stem~\cite{rouard2025musicgen}, which integrates both functions, has been recently released. However, it was released during our study and supports only three stems. While recent studies provide additional controls for TTM models~\cite{wu2024music,tal2024joint}, text instructions continue to be the dominant approach. Given this and the study's limited duration, we decided to focus the study on models that use text as input. Hence, we do not use the melody conditioning of MusicGen.  

\textbf{Interface design}.
\begin{figure*}[t!]
    \centering
    \subfloat[]{
        \includegraphics[width=.48\linewidth]{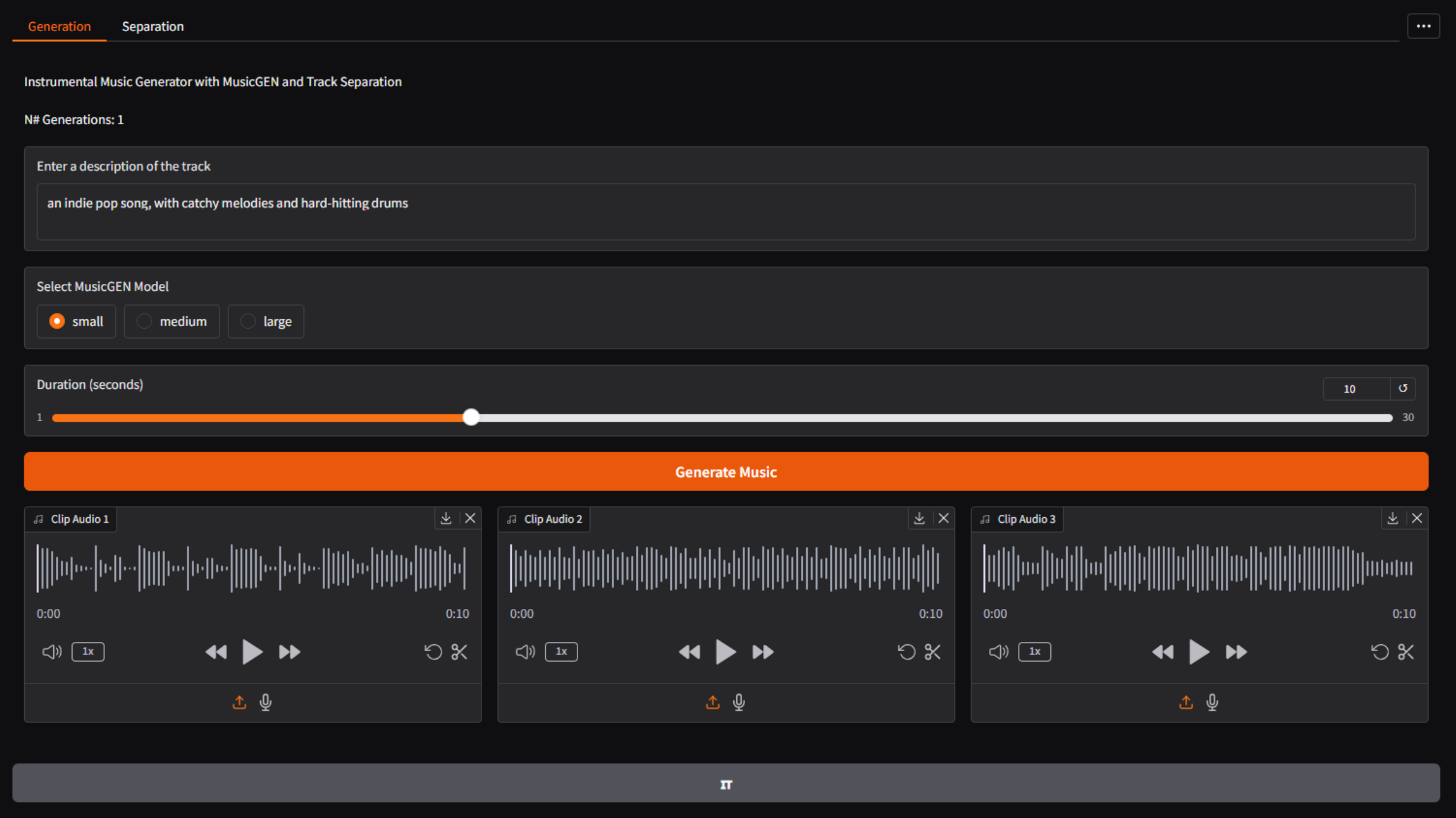}
    }
    \subfloat[]{
        \includegraphics[width=0.48\linewidth]{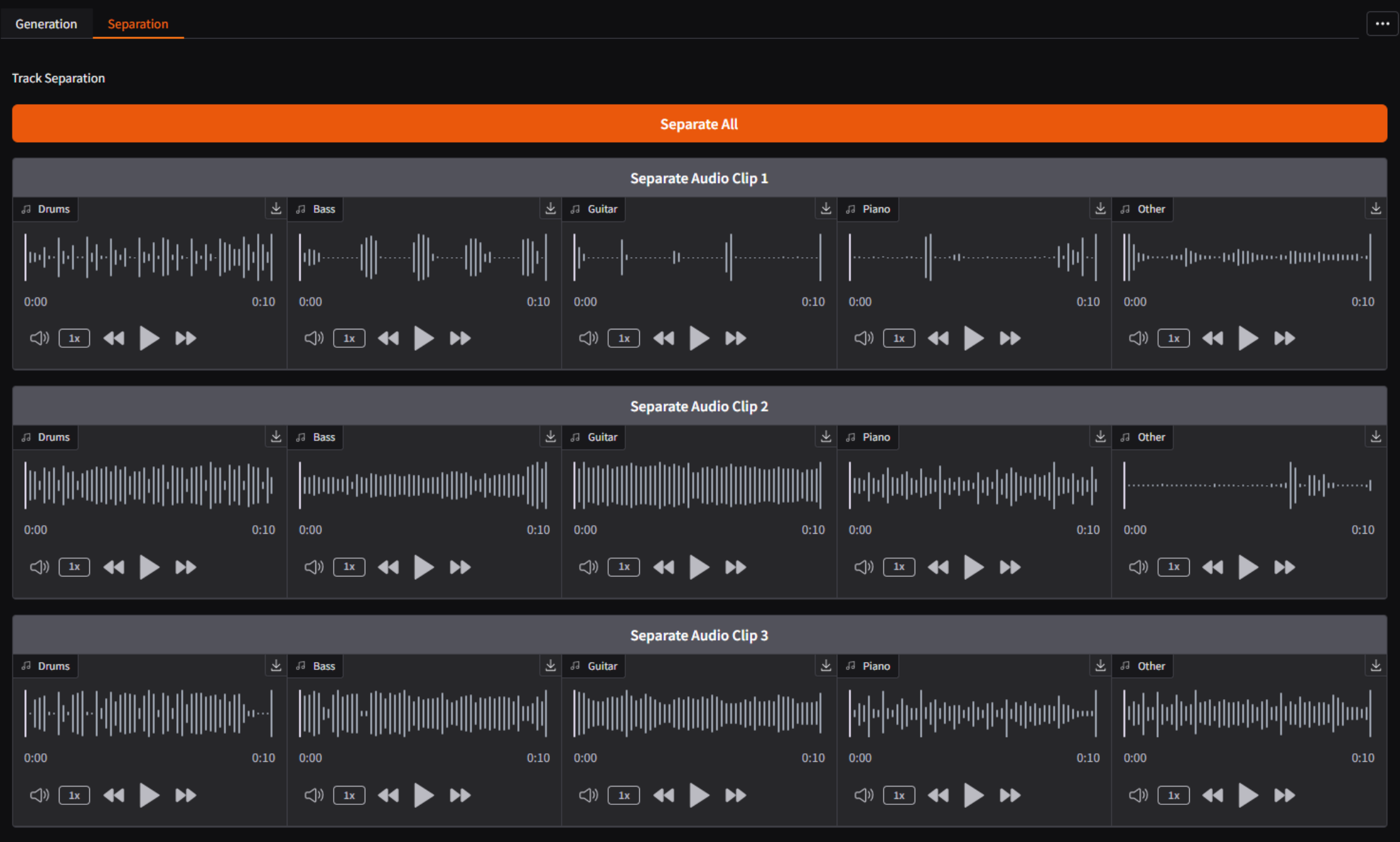}
    }
    \caption{User interface: (a) music generation  model; (b) source separation model.}
    \label{fig:interface}
\end{figure*}  
We designed the interface based on the \textit{AI controllability exploration principles} proposed in~\cite{weisz2023toward}. Inspired by~\cite{ronchini2024paguri,kamath2024sound}, we implemented \textit{multiple output principles} together with \textit{general} and \textit{domain-specific} controls, allowing users to explore a wide range of possibilities and better understand the model's capabilities. To decide which control options to include, we analyzed some interfaces already available on Hugging Face. To ensure ease of use and accommodate the experiment's time constraints, we kept the interface simple with a limited set of controls. 
The interface, shown in Figure~\ref{fig:interface}, presents one tab for music generation and one tab for source separation. Users can control the generation by selecting output duration, MusicGen size, and input prompt. The separation interface offers stems for \textit{drums, bass, guitar, piano}, and \textit{other}, with vocals excluded as MusicGen does not generate them. It was developed with Gradio for its intuitive design, rapid deployment, and focus on machine learning models. We built a custom interface for better experiment control; one author developed it, and two tested it.

\textbf{Experimental Procedure}. The experiment was conducted live, either in person or online, using the same procedure. Participants scheduled a one-hour session with the experimenter and used their own laptops to use their preferred setup (e.g. DAW, headphones) to produce music. They accessed the interface through a web app, facilitated by ngrok~\cite{ngrok}, 
 with secure remote access via SSH. The interface ran on a Quadro P6000 GPU on a private cluster. Each session was divided into three steps:

\textit{1. Pre-experiment step}: participants were introduced to the experiment and then completed a survey on their music production experience, familiarity with AI tools, and expectations in the creative workflow.

\textit{2. Hands-on step}: participants were invited to create a music excerpt using the interface and their preferred setup. They prompted the model, using the available controls, then generated and downloaded the audio or used the separation model to obtain individual stems from the generated audio. They had full freedom to interact with the models, and restart the generation process as desired, with no music genre limitations. 
The session was recorded via screen and audio, with observational notes taken throughout.

\textit{3. Post-experiment step}: at the end of the session, participants evaluated their user experience, workflow creativity, productivity, TTM effectiveness, ethical considerations, and provided additional feedback in a semi-structured interview. 

The pre- and post-session questionnaires were adapted from~\cite{newman2023human,ronchini2024paguri,morreale2016collaborating} and the Goldsmiths Musical Sophistication Index~\cite{mullensiefen2014musicality}. Participants were recruited through music producer communities and relevant mailing lists, with a minimum level of music production experience required. A total of $17$ international participants from 7 different countries participated in the study ($8$ from Italy, $2$ from India, $2$ from the U.S.A, $2$ from the People's Republic of China, and $1$ each from the Republic of Korea, Greece, and Spain). $4$ participants attended in person and $13$ remotely via Zoom platform; each participant was assigned a random ID.

\section{Results}
\label{sec:results}

Section~\ref{subsec:musicexperince} and section~\ref{subsec:musiccreativity} present Likert-scale and multiple/single-choice answers from the pre- and post-experiment questionnaire, respectively. Section~\ref{subsec:TA} provides a thematic analysis of the open-ended questions of the post-questionnaire.
Likert-scale answers are shown using diverging plots, where positive responses are on the right, negative on the left, and neutral responses are split between both sides. The bar length represents the number of responses, 
different colors 
indicate specific scores. For the Likert-scale-based questions, we report the ratings on a scale from 1 to 5 due to space limitations. However, in the questionnaire, each question retained its appropriate responses according to the Likert-scale answer guide\footnote{\href{https://www.extension.iastate.edu/documents/anr/likertscaleexamplesforsurveys.pdf}{www.extension.iastate.edu/documents/anr/likertscaleexamplesforsurveys.pdf}}.

\subsection{Pre-experiment questionnaire: Musical background and AI tool familiarity}
\label{subsec:musicexperince}
\begin{figure*}[t!]
  \centering
  \includegraphics[alt={Diverging bar charts showing users' familiarity with AI tools and users' expectation for AI to improve creativity.}, width=1.\linewidth]{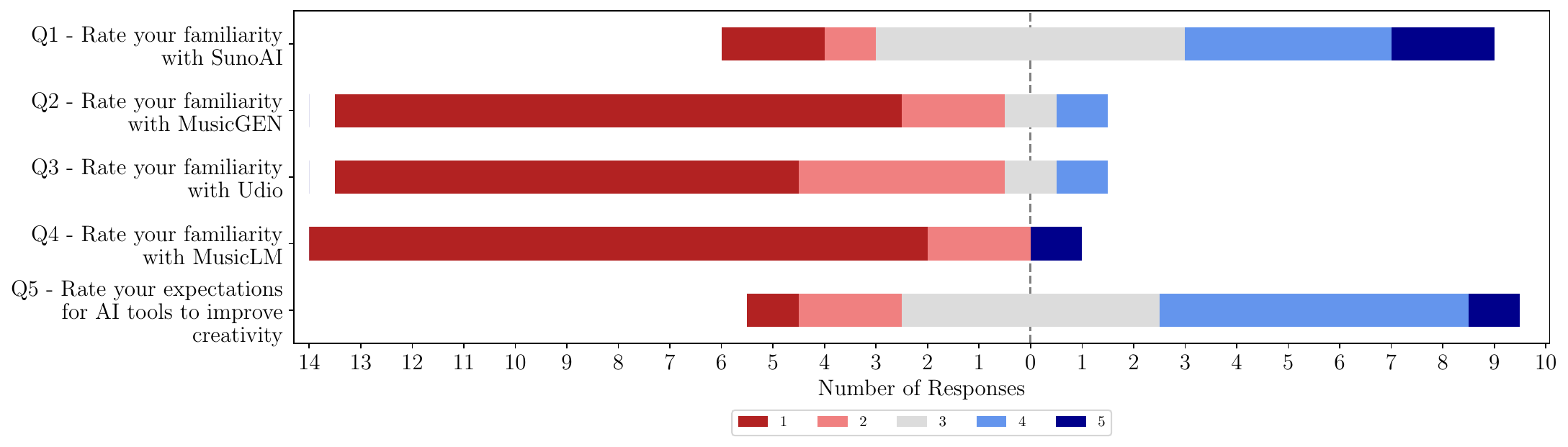}
  \caption{Diverging bar charts showing users' familiarity with AI tools and users' expectation for AI to improve creativity.}
  \label{fig:famAItools}
\end{figure*}

\textbf{Music Production Experience and Genre}. Among participants, $10.5\%$ have more than 5 years of experience, $42.1\%$ have 3-5 years, $21.1\%$ have 1-3 years, and $26.3\%$ have less than 1 year of experience. Logic Pro is the most popular DAW, used by $35.7\%$ of participants, followed by Ableton at $28.6\%$ and Reaper at $14.3\%$. Other DAWs mentioned included Studio One, Cubase, Pro Tools, and Samplitude. $76.5\%$ of participants had used AI tools for music creation before. 
The main genre of music production is \textit{Pop}, mentioned $8$ times, then \textit{Classical}, \textit{House / EDM / Dubstep} and \textit{Rap / Trap} mentioned $5$ times each, while \textit{Rock / Metal} and \textit{R\&B} and \textit{Lo-Fi} were mentioned $3$ times each. \textit{Carnatic Music, Urban, Latin, and Acoustic} appeared once each. They could list multiple genres. 

\textbf{TTM models familiarity and expectation}. Fig.~\ref{fig:famAItools} shows users' familiarity with key music generative models and their expectations for AI tools to enhance music creativity. Other mentioned AI tools include SynthPlant, NetEase, Tianyin AI, Watson Beats, ImprovNet, Stable Audio, Beatoven.ai, Soundraw, Mubert, iLoveSong.ai, and AIVA. Suno AI stands out, likely due to stronger brand recognition or better accessibility compared to others. While there is an optimistic outlook on AI’s role in creative processes, some participants remain cautious. To gain deeper insight into these expectations, we asked participants what they hoped to achieve using AI tools in their creative workflow. A brief thematic analysis  of their answers revealed key themes: efficiency and time savings, inspiration, customization and personalization, simplification for less experienced producers, skill development, and sound quality. 

\subsection{Post-experiment questionnaire: AI tools for music productivity}
\label{subsec:musiccreativity}
\begin{figure*}[t!]
  \centering
  \includegraphics[alt={Likert-scale answers from the post-experiment questionnaire.},width=1.\linewidth]{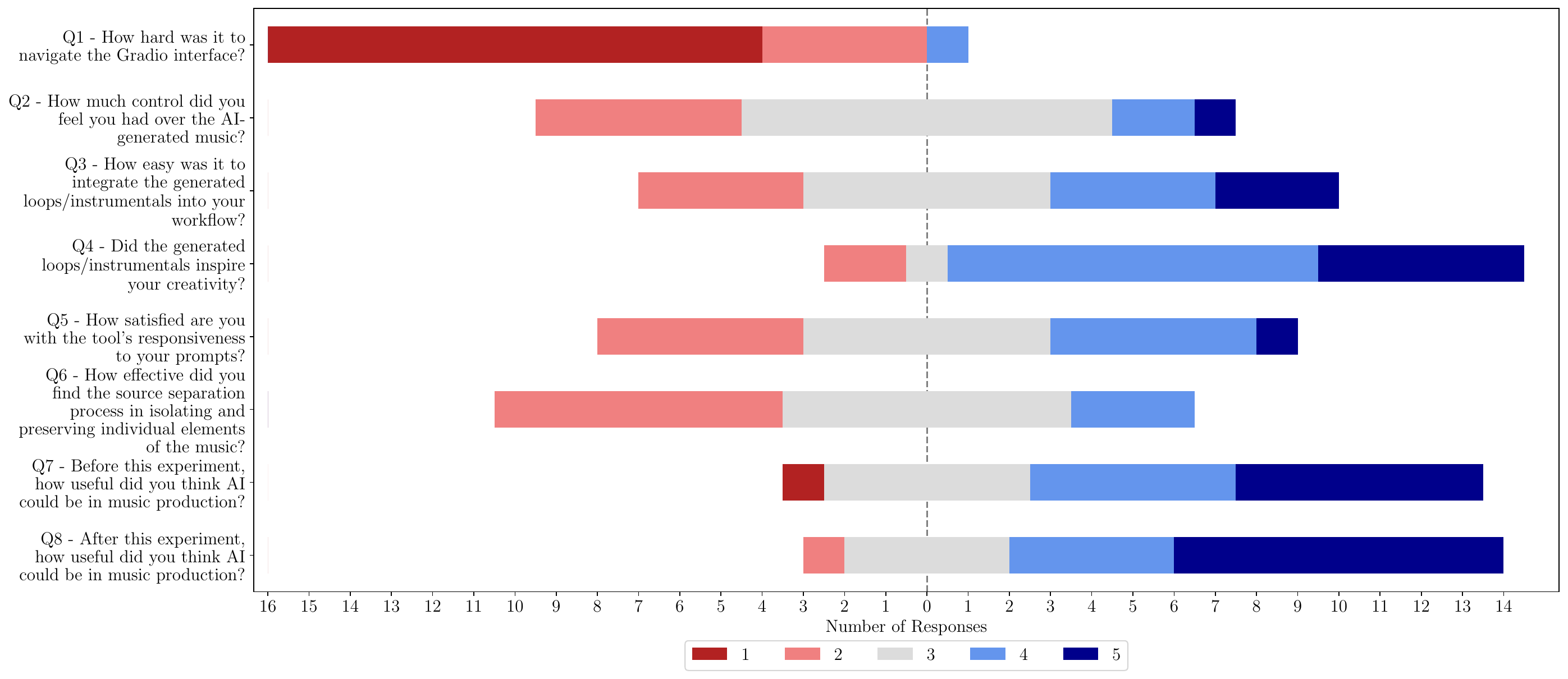}
  \caption{Likert-scale answers from the post-experiment questionnaire.}
  \label{fig:AItools}
\end{figure*} 
\textbf{Likert-scale questions}. Figure~\ref{fig:AItools} shows answers to 5-points Likert-scale questions. 
Participants responded positively to integrating AI-generated loops into their workflow, with many finding the loops inspiring creativity, positioning the tool as a potential creative collaborator. Perceptions of control over AI outputs and satisfaction with the tool’s responsiveness varied. Opinions on the source separation feature were similarly mixed. 

\textbf{Single and multiple-choices questions}. We asked participants where they would integrate the tool in the music production workflow. $94.12\%$ of participants would use it \textit{``during the ideation phase''}, while $47.06\%$ consider it potential for \textit{``variations and experimentation''}; $35.29\%$ identified composition and arrangement as a key stage, and $23.53\%$ mentioned \textit{``remixing and reworking existing tracks.''} Other responses included \textit{``sound and track generation''} and \textit{``digital audio generation''}. 
We also asked if they would incorporate an AI-based source separation tool into their production workflow. $52.94\%$ expressed a strong interest in using it for music production, while $41.18\%$ indicated they would use it only in specific scenarios. $5.88\%$ preferred manual mixing or alternative methods. We then explored which applications participants found most suitable for AI-generated music. $88.24\%$ identified \textit{``background music for games, YouTube videos, or other media''} and \textit{``music production assistance''} as primary use cases, highlighting its potential to support the creative process. $64.71\%$ pointed to \textit{``experimental production and sound design''}, reflecting an interest in exploring new creative directions. $47.06\%$ consider it useful for \textit{``film, TV, or commercial soundtracks,''} while $35.29\%$ mentioned \textit{``remixing and reinterpreting existing tracks''}. $17.65\%$ would use it for \textit{``direct listening''}. 

\subsection{Thematic analysis}
\label{subsec:TA}

An inductive thematic analysis was performed on the open-ended questions, resulting in $39$ initial codes, which were iteratively merged into $14$ themes, organized into $3$ sections using affinity diagrams~\cite{kamath2024sound}. The sections are: productivity and effectiveness, usability and creative workflows, and ethical considerations. One author drafted the initial codebook, while two other authors iteratively refined the final themes, which are reported in bold in the following subsections.

\subsubsection{Productivity and effectiveness.}

\textbf{\textit{Creative Misalignment}}. Discrepancy between participants' creative visions and the generated output emerged as a key theme. P12: \textit{``It didn't go well with my initial ideas as planned''}, P14: \textit{``I had something in mind but it proposed me some different and unexpected things''}, highlighting a mismatch between expectations and results. P15 mentioned, referring to the content generated: \textit{``sometimes it deviates from what I had in mind''}, emphasizing the challenge of aligning with personal projects. P16: \textit{``I had some creative ideas about how the song I wanted should sound like, but I didn't get the result that I had in mind''}, while P2 stated: \textit{``The content generated didn't reflect what I had in mind''}. P11: \textit{``Some of the instrument sounds I asked for didn't come out as expected''}, further illustrating this gap. 

\textbf{\textit{Integration Challenges.}} Tempo, key, and beat alignment emerged as main issues for integrating generated content into projects. P12: \textit{``Generated output was not loop-based, and the tempo was not accurate. It was difficult to slice them and beat-match to another track. About key, I wasn't sure if they were correct''}. P2 similarly pointed out (referring to the fact that the content generated did not reflect what they had in mind): \textit{``[..], especially in terms of BPM''}. Source separation also posed difficulties. P17: \textit{``The guitar and piano separation tools did not pick up audible frequencies''}, 
P8 noted: \textit{``Noisy samples after the separation''}. Participants P8 and P10 faced challenges with one-shot samples generation.

\textbf{\textit{From expectation to innovation}}. When outputs did not align with their original ideas, many participants changed their approach. P2: \textit{``I can't use the generated samples as main elements, but they’re useful and beautiful as background, not what I originally intended''}.  
P7: \textit{``The AI tool redefined my project, inspiring new ways to create rather than helping me achieve my original idea''}. Similarly, P12 added, referring to the fact that it did not go well with their initial idea: \textit{``However, by modifying the samples in GarageBand’s audio-flex functions, I was able to discover a very new direction of experimental music''}. P14: \textit{``It transformed my idea into something new. I had something in mind, but it proposed different and unexpected things''}. These responses reflect how, despite results not aligning with initial ideas, the TTM model became a source of unexpected inspiration, encouraging participants to 
explore new creative directions.

\textbf{\textit{AI as an efficient collaborator}}. Some participants considered the tool as an efficient collaborator and a creative partner. P4 highlighted the tool's usefulness, even if \textit{``just for accelerating the generation process of something very precise I have in mind''}. P11: \textit{``In some cases, it's easier to generate what I want than it is to search for a loop''}. P17: \textit{``It created a good foundation for work to continue but gave space for me to build something of my own, which is nice, not too prescriptive''}. These insights suggest that, in specific scenarios, AI can speed-up the process and be an efficient collaborator.
\subsubsection{Usability and creative workflows.} 
\textbf{\textit{Need for greater control}}. Concerns were raised about limitations in controlling the generation process. P15 recommended integrating \textit{``multi-modal prompts''} for versatility. Many participants highlighted the need for control over tempo alignment, key selection, and loop duration. P17 proposed \textit{``More parameters like key signature or BPM for better alignment of the output''}. P8 emphasized the need of \textit{``enhanced controllability of accurate BPM and key info''}, while P6 wanted to \textit{“change/select tempo and measure of the loop”}. P16 recommended \textit{“fine-grained control''}, while P1 suggested to allow users to choose the type of input prompt (text, tag, or reference audio). 

\textbf{\textit{Lack of Editability}}. Participants expressed frustration with the inability to modify or refine the generated content. P6: \textit{``It was frustrating that I was not able to modify the tracks/ideas given"}. P7 suggested adding \textit{``a second prompt to a result, memorizing it and changing only a part of it''} for more control over specific elements. P12 emphasized the need for an \textit{``iterative process of re-generation of samples''} and proposed \textit{``music editing through a piano-roll interface on the generated audio samples''} for precise manipulation. 

\textbf{\textit{Partial usability of generated content.}} Concerns were raised about the limited usability of the generated content. P11: \textit{``After the clips were generated, some would be usable and others wouldn't. Also, some of the instrument sounds I asked for didn't come out as expected''}, highlighting that only certain parts of the output met expectations. P12 mentioned: \textit{``I expected a stem-wise output (e.g., by specifying jazz-marimba-solo), but it often comes with other instruments mixed in''}, emphasizing the challenge of isolating specific sounds.  

\subsubsection{Ethical considerations.}
\label{subsec:ethics}

\textbf{\textit{Copyright concerns and AI regulation}}. Copyright regulations were among the main considerations. P3: \textit{``a group of experts should intervene to evaluate how much an AI-generated track takes from existing pieces''}, suggesting the need for human oversight. P5 emphasized the importance of \textit{``strict regulations about copyright and AI training data''}. P17 called for a \textit{``broader overhaul of the copyright system''}, advocating for a systemic shift in how copyright laws should adapt to AI-generated works.

\textbf{\textit{Artist Compensation}}. P17: \textit{``revision of the way artists are compensated for their work''}. P16 argued that \textit{``artists whose work has been used to train AI should be compensated''}. P4 stressed the importance of ensuring that \textit{``artists are compensated each time their data is used''}. 

\textbf{\textit{Diversity and risk of homogenization in AI-generated music}}. The risk of homogenizing music was noted, stressing the need for diversity in music generation. P17: \textit{``AI training should span a wide range of music styles''}; P3: \textit{``we should limit ourselves to get inspiration from the AI and to ease the sound generation process, not actually to make the AI create our music''}. P6 pointed out that in generating ethnic music genres, the model \textit{``fell short''}, underscoring the need for better representation of diverse cultural sounds. 

\section{Discussion}
\label{sec:discussion}

While TTM models show promise in artistic applications, they primarily serve as aids for sketching and inspiration rather than as production-ready solutions. This reinforces the idea that AI serves as an innovative set of tools for creative exploration~\cite{icsik2024exploring}. However, integrating TTM-generated audio samples into human creative workflows remains challenging due to control limitations. 
We recommend practical design improvements that address current limitations, such as improving tempo, key, and beat controls, editing capabilities, support for iterative re-generation, and, very importantly, DAW integration. These features would greatly strengthen the usability and creative potential of the tools.
Additionally, interpretative gaps often arise between the AI’s output and the artist’s intended vision, particularly when producers have a clear idea of what they want. Some producers embrace these tools as a source of inspiration, adapting their creative process to incorporate the AI’s output. Others, however, report having to modify (sometimes completely changing) their original idea to accommodate the model’s unexpected outputs. This dynamic illuminates the role of AI as inspiration, prompting a question: does AI empower artists by enhancing creative exploration, or, on the flip side, does it constrain the process by steering creativity in unintended directions? Similar patterns are observed in~\cite{fu2025exploring}. 

Creative ideas are often discarded when they fail to integrate with other creative elements. Future models should consider additional input controls and allow for editable content to better align outputs with users' intentions and empower them to refine specific sections. These challenges were already highlighted in earlier research, which led to the introduction of controls for tempo, rhythm, and chord progression~\cite{wu2024music,tal2024joint,lan2024musicongen,colombo2025mambafoley}. However, many models are primarily evaluated based on technical performance, such as audio quality or prompt adherence, rather than through user-centered case studies. Misalignment between user intent and generated output persists, as evidenced in~\cite{fu2025exploring}, which evaluated several AI tools. This emphasizes the need for systems that support iterative refinement, giving artists greater creative control. Previous work~\cite{zhang2024musicmagus} has explored such approaches, offering a promising direction for future research. Additionally, personalization features~\cite{ronchini2024paguri} and source separation techniques, as proposed in this study and by~\cite{rouard2025musicgen}, show promise in addressing these gaps. 

Copyright regulation, artist compensation, and content diversity emerge as key concerns from this study. Previous research~\cite{sarmento2024between,ronchini2024paguri} underscores the importance of diverse datasets to avoid cultural biases and promote a wider range of artistic expression. This study reveals how the limitation of generating sub-genres and less-represented music has raised concerns among music producers about homogenization, raising another question: are AI tools encouraging creativity and diversity, or are they contributing to the homogenization of music genres and creation? To address these issues, we argue that evaluation methods for TTM models must evolve beyond traditional listening tests and technical performance. Adopting human-centered evaluation methods (with direct interaction and feedback) is a more accurate way of assessing AI’s effectiveness in creative workflows. Hands-on testing by artists is essential for understanding whether these systems enhance or hinder their creative processes. Since music creation relies heavily on human intuition and iterative refinement, transitioning AI from a passive tool to an active, empowering creative partner requires more control, diverse datasets, and user-centric evaluation. We believe it is equally important to understand user interactions with these models~\cite{gui2024adapting}, and prioritize human-centered development to address real-world usability~\cite{ronchini2024paguri,zang2024interpretation,messina2025mitigating}.
This research highlights the importance of ethical AI development and active collaboration with musicians to ensure that AI enhances artistic creativity rather than undermining human expression.

\section{Conclusions}
\label{sec:conclusion}

This study explores the interaction of experienced music producers with an interface combining a Text-to-Music model and an AI-based music source separation tool to create a music track. Feedback was gathered through semi-structured interviews. The results show that TTM models inspire ideas but struggle to integrate them with human-made music, especially when producers have specific intentions. This often leads to a shift in approach in using the models, raising questions about their role as creative partners and whether human-centered testing should be a key metric for evaluating their effectiveness. More control over the generation process is needed. The risk of homogenization from Western-centric training data, artist compensation, and copyright concerns also emerged.

\begin{credits}
\subsubsection{\ackname} The authors acknowledge support from the IEEE Signal Processing Society under the Signal Processing Society Scholarship Program.

\subsubsection{\discintname}
The authors have no competing interests to declare that are
relevant to the content of this article. 
\end{credits}

\bibliographystyle{splncs04}
\bibliography{cmmr2025_camera-ready_template}
\end{document}